# DNA-decorated carbon nanotubes for chemical sensing

Cristian Staii[1], Michelle Chen[2], Alan Gelperin[3], and Alan T. Johnson, Jr. [1,*]

[1]*Department of Physics and Astronomy and Laboratory for Research on the Structure of Matter, University of Pennsylvania, Philadelphia, Pennsylvania 19104*

[2]*Department of Material Science and Engineering, University of Pennsylvania, Philadelphia, Pennsylvania 19104*

[3]*Monell Chemical Senses Center, Philadelphia, Pennsylvania 19104*

[*]Corresponding author. Email: cjohnson@physics.upenn.edu

Dated (September 1, 2005)

**Abstract**

We demonstrate a new, versatile class of nanoscale chemical sensors based on single-stranded DNA (ss-DNA) as the chemical recognition site and single-walled carbon nanotube field effect transistors (swCN-FETs) as the electronic read-out component. SwCN-FETs with a nanoscale coating of ss-DNA respond to gas odors that do not cause a detectable conductivity change in bare devices. Responses of ss-DNA/swCN-FETs differ in sign and magnitude for different gases, and can be tuned by choosing the *base sequence* of the ss-DNA. Ss-DNA/swCN-FET sensors detect a variety of odors, with rapid response and recovery times on the scale of seconds. The sensor surface is self-regenerating: samples maintain a constant response with no need for sensor refreshing through at least 50 gas exposure cycles. This remarkable set of attributes makes sensors based on ss-DNA decorated nanotubes very promising for "electronic nose" and "electronic tongue" applications ranging from homeland security to disease diagnosis.

The one-dimensional carbon cage structure of semiconducting single-walled carbon nanotubes (swCNs) makes their physical properties exquisitely sensitive to variations in the surrounding electrostatic environment, whether the swCNs are suspended in liquid or incorporated into field effect transistor (FET) circuits on a substrate.[1-3] Bare and polymer-coated swCNs are reported to be sensitive to various gases,[4-9] but swCNs functionalized with biomolecular complexes hold great promise as molecular probes and sensors[10-13] targeted for chemical species that interact weakly or not at all with unmodified nanotubes. Derivatized swCN-FETs are attractive as electronic-readout molecular sensors due to their high sensitivity, fast response time, and compatibility with dense array fabrication.[3] Derivatized semiconductor nanowires have similar performance advantages, and recent work indicates they are also promising candidates for gas[14] and liquid-phase sensors.[15,16]

An effective scheme to functionalize swCN-FET sensors should simultaneously achieve robust, reproducible decoration of the swCN with molecular flexibility promising sensitivity to a wide spectrum of analytes. Non-covalent functionalization is required to avoid degrading the high-quality electronic properties of the swCN-FET.

Nucleic acid biopolymers are intriguing candidates for the molecular targeting layer since they can be *engineered*, using directed evolution, for affinity to a wide variety of targets, including small molecules and specific proteins.[17,18] High throughput screening of multiple oligomers was used recently to select films of dye-labeled ss-DNA oligomers that function as gas sensors.[19,20] On exposure to an odor, fluorescence of the intercalated dye changes relative to the level measured when the sample is exposed to clean air. The molecular mechanism of this response is not known, but the response to particular odors was reported to be specific for the base sequence of the oligomer. Ss-DNA is also known to have high affinity for swCNs due to a favorable $\pi$-$\pi$ stacking interaction.[21]

These findings motivated our exploration of the ss-DNA/swCN-FET hybrid nanostructure as a gas sensor with electronic readout. We focus here on devices consisting of individual nanotubes contacted by electrodes in order to illuminate intrinsic properties of the ss-DNA/swCN system. Sensors based on swCN networks may be easier to manufacture in functional systems, but they introduce complicating, poorly controlled factors, e.g. the presence of both metallic and semiconducting swCNs, and tube-tube cross junctions.

SwCNs were grown by catalytic chemical vapor deposition (CVD) on a $SiO_2$/Si substrate. FET circuits were fabricated with Cr/Au source and drain electrodes patterned using electron beam lithography and the degenerately doped silicon substrate used as a backgate (Figure 1).[22] For each device, source-drain current I was measured as a function of bias voltage $V_B$ and gate voltage $V_G$ under ambient laboratory conditions. Circuits consisting of individual p-type semiconducting nanotubes, where the carriers are positively charged holes, were selected by using only devices that showed a strong decrease in $I(V_G)$ for positive $V_G$ (ON/OFF ratio exceeding 1000).

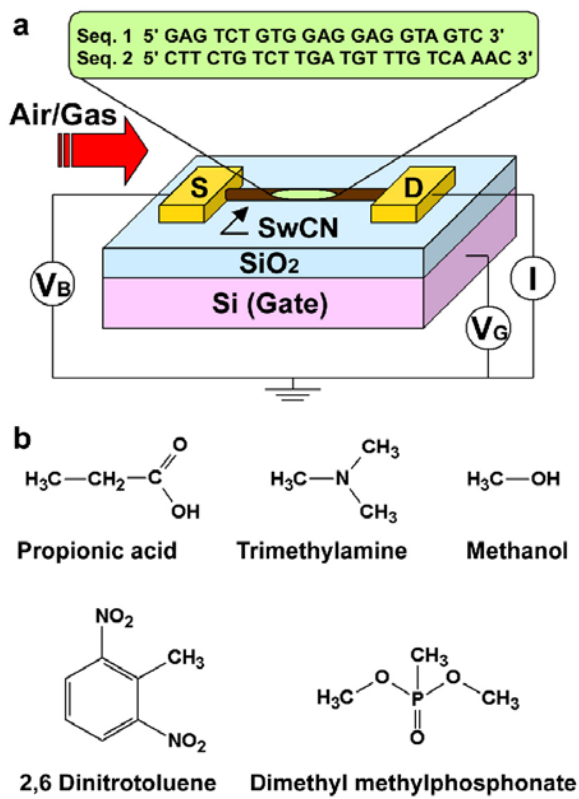

**Figure 1.** (a) Schematic of the experimental setup. (b) Gases (odors) used in the experiment.

Two ss-DNA sequences were chosen based on prior work:[19,20]
    Sequence 1  5' GAG TCT GTG GAG GAG GTA GTC 3'
    Sequence 2 5' CTT CTG TCT TGA TGT TTG TCA AAC 3'
Oligonucleotides were obtained from Invitrogen (Carlsbad, CA) and diluted in distilled water to make a stock solution of 658 µg/ml (Seq. 1) or 728 µg/ml (Seq. 2). After odor responses of the bare swCN-FET device were measured, a 500 µm diameter drop of ss-DNA solution was applied to the device for 45 min, and then dried in a nitrogen stream. About 25 devices from two different swCN growth runs were selected for detailed analysis and treated with ss-DNA for the experiments described here. Statistical analysis of atomic force microscopy (AFM) images of the same tube before and after DNA application showed an increase in the nominal tube diameter from 5.4 ± 0.1 nm to 7.2 ± 0.2 nm, indicating formation of a nanoscale layer of ss-DNA on the swCN surface (Figure 2a, b). For both of the sequences used, application of ss-DNA caused the threshold value of $V_G$ for measurable conduction to decrease by 3 – 4 V (Figure 2c). This corresponds to a hole density decrease of roughly 400/µm, assuming a backgate capacitance (25 aF/µm) that is typical for this device geometry.[22] Furthermore, the "ON" state conductivity of the ss-DNA/swCN-FET was ~ 10% lower than that of the bare device (Figure 2c), suggesting weak carrier scattering by the molecular coating.

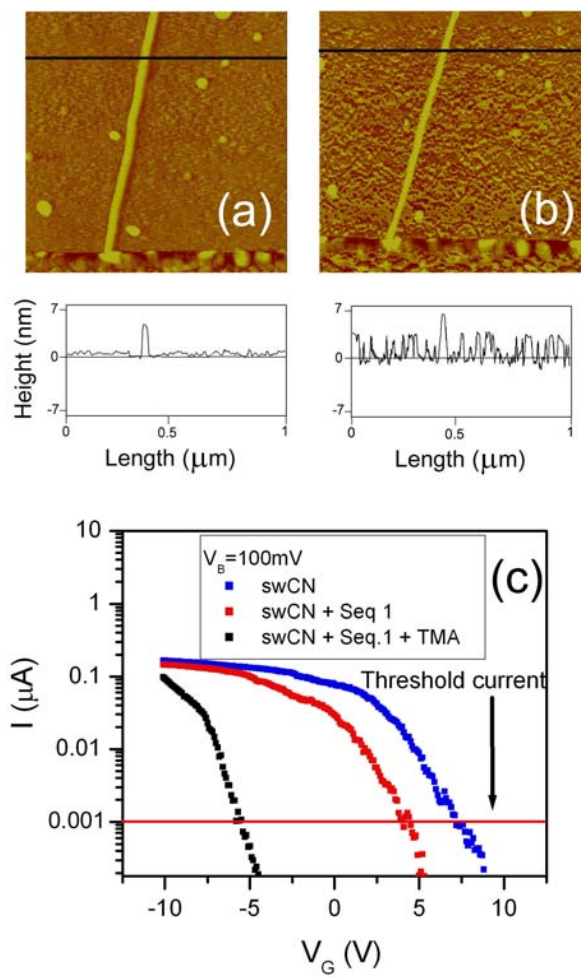

**Figure 2.** (a), (b) AFM images (1 μm x 1 μm, z-range 10 nm) and line scans of the *same* swCN before (a) and after (b) functionalization with ss-DNA. The measured diameter of the bare swCN is 5.4 ± 0.1 nm, while after application of ss-DNA its diameter is 7.2 ± 0.2 nm. The increase in surface roughness in (b) is attributed to non-specific binding of ss-DNA to the $SiO_2$ substrate. (c) Current (I) versus backgate voltage ($V_G$) characteristic of a bare swCN-FET sensor (blue), the same device after functionalization with ss-DNA sequence 1 and exposed to air, the same ss-DNA/swCN-FET exposed to TMA vapor. Source-drain bias voltage is 100 mV.

Figure 1(b) shows the five odors used to characterize the sensor response: methanol, propionic acid, trimethylamine (TMA), dinitrotoluene (DNT), and dimethyl methylphosphonate (DMMP; a simulant for the nerve agent sarin[23]). For DNT, a liquid solution was prepared by dissolving 50 mg/mL of the material in dipropylene glycol.

A reservoir of saturated vapor of each odor was prepared and connected to a peristaltic pump and switching valve array so the flow of room air directed over the device (0.1 ml/sec) could be electrically diverted to one of the odor reservoirs for a set time (typically 50 sec), after which the flow reverted to plain air. The air or air/analyte mixture was directed towards the sample through a 2 ± 0.1 mm-diameter nozzle positioned 6 ± 1 mm above the sample surface. For each analyte, we estimate the concentration delivered to the sample to be 3 % of the appropriate saturated vapor pressure (see Table 1). The source-drain current (I) through the device was measured as a function of gate voltage $V_G$ for a fixed bias voltage $V_B$. For each sample (both before and after application of ss-DNA), it was found that $V_G = 0$ V was a region of large transconductance ($dI/dV_G$), indicating high sensitivity of the swCN-FET to environmental perturbations. Detailed measurements of the odor-induced changes in I as a function of $V_G$ and $V_B$ and sensor response as a function of analyte concentration will be the subject of future work. Here we focus on odor-induced changes in the current measured with $V_B = 100$ mV and $V_G = 0$ V.

In Figure 3a,b, we show responses of two devices to odors, before (blue points) and after (red points) coating with ss-DNA Seq. 2. The current response of a bare swCN-FET was less than our experimental sensitivity ($\Delta I/I \sim 1\%$) when exposed to methanol (Figure 3a), propionic acid, DMMP and DNT (data not shown). After coating this same device with ss-DNA Seq. 2, exposure to methanol gives a 20% decrease in the transport current. We conclude that the ss-DNA layer increases the binding affinity for methanol to the device, thereby increasing the sensor response. Even when a bare swCN sensor responds to a particular gas ($\Delta I/I = -10\%$ on exposure to TMA, Fig. 3b), functionalization with ss-DNA enhances the molecular affinity and associated response ($\Delta I/I = -30\%$).

We observe further that different odors elicit different current responses from ss-DNA/swCN-FET sensors. For example, the response to propionic acid of a device with ss-DNA Seq 1 differs in both sign and magnitude from the response to methanol (Figure 4c). The data in Figure 4 also demonstrate a constant sensor response is maintained through multiple odor exposures. As a test of response reproducibility, we exposed a device to 50 cycles of TMA and air exposure (odor and air pulses each 50 seconds in duration), and the response was maintained to within 5% (data not shown). Device-to-device variation in odor response is also small (see Table 1 and discussion below). This excellent reproducibility for a single device and across devices indicates very favorable prospects for quantitative modeling of individual devices and integrated systems.[24]

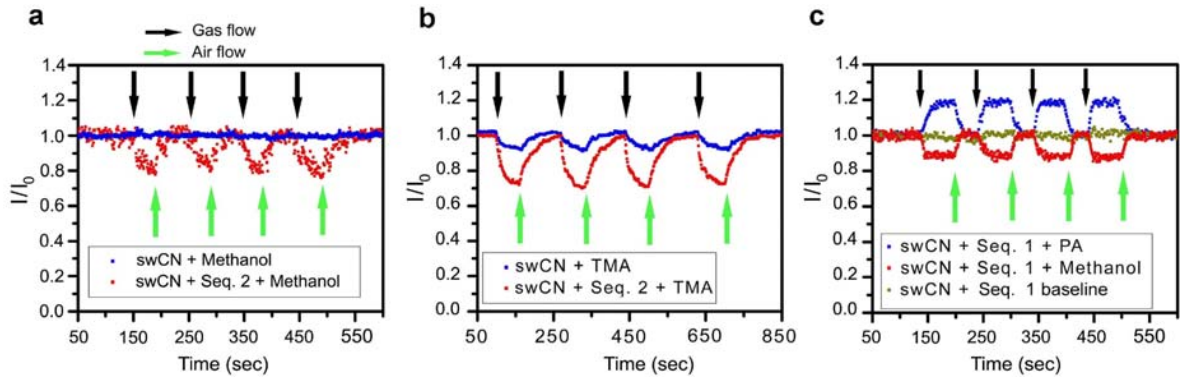

**Figure 3.** Change in sensor current upon odor exposure. Currents are normalized to $I_0$, the value when exposed to air (no odor). (a) Bare swCN-FET does not respond to methanol vapor (blue points). The same device coated with ss-DNA sequence 2 (Seq. 2) shows clear responses to methanol (red points). (b) A second bare device responds to TMA (blue points) but after application of Seq. 2, the response is tripled (red points). (c) The sensor response to propionic acid (blue points) differs in sign and magnitude from the response to methanol (red points). Green data are the current baseline (no odor). $V_B$ = 100 mV and $V_G$ = 0 V for all data sets.

Finally we find that the odor response characteristics of ss-DNA/swCN-FET sensors are specific to the *base sequence* of the ss-DNA used (Table 1). The number of distinct ss-DNA 24-mers is extremely large, and they are all expected to bind readily to swCNs through a π−π stacking interaction. It should be possible to create a large family of sensors with disparate odor response characteristics, an important building block of "electronic nose" and "electronic tongue" systems discussed below.[25]

| Odor | vapor pressure (Torr) | Est. conc. (ppm) | bare swCN %ΔI/I | swCN + Seq. 1 %ΔI/I | swCN + Seq. 2 %ΔI/I |
|---|---|---|---|---|---|
| Water | 17.5 | 700 | 0 ± 1 | 0 ± 1 | 0 ± 1 |
| Propionic acid | 4 | 150 | 0 ± 1 | +17 ± 2 | +8 ± 1 |
| TMA | 500 | 20000 | -9 ± 2 | -20 ± 2 | -30 ± 2 |
| Methanol | 100 | 4000 | 0 ± 1 | -12 ± 2 | -20 ± 2 |
| DMMP | 0.6 | 25 | 0 ± 1 | -14 ± 2 | -7 ± 2 |
| DNT | 1 | 40 | 0 ± 1 | -14 ± 4 | -4 ± 2 |

**Table 1**. Measured responses of devices to gaseous analytes. Estimated concentration corresponds to 3% of the saturation vapor pressure (www.sciencestuff.com). Each quoted sensor response is based on measurements of 5-10 different devices. Uncertainties are the standard deviation of the mean.

To explore this possibility, we measured the odor response of ss-DNA/swCN-FET sensors to DNT and DMMP, simulants for explosive vapor and nerve gas, respectively. As seen in Figure 4 and Table 1, ss-DNA functionalized swCN-FETs respond to these two odors while bare devices do not, and the response characteristic is specific to the ss-DNA sequence used to decorate the device. Control experiments were conducted to verify that the ss-DNA/swCN-FET sensor showed no response to dipropylene glycol, the solvent used for DNT, and water vapor, a common background substance. The signal-to-noise levels of the measurements in Figure 4 indicate that detection of concentrations less than 1 ppm should be possible even with these unoptimized devices. The DMMP concentration used in the experiment is estimated to be 25 ppm, and the observed response is distinct but modest (ΔI/I ∼ -7% for ss-DNA sequence 1 and –14 % for ss-DNA sequence 2). Thin-film transistor sensors fabricated from swCN *networks* are reported to respond much more strongly to DMMP (ΔI/I ∼ -50 % at concentration of 1 ppb[6]). Our experiment indicates this large response is not intrinsic to individual swCNs but may be related to the network connectivity.

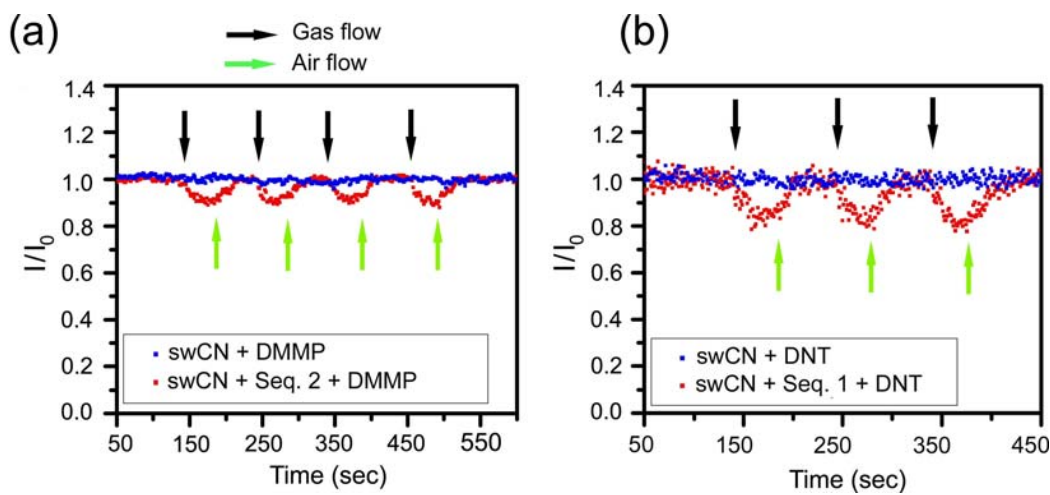

**Figure 5**. (a) Change in the device current when sarin-simulant DMMP is applied to swCN-FETs before and after ss-DNA functionalization. (b) Sensor response to DNT.

We briefly consider the mechanism of molecular detection, recognizing that many important issues remain to be clarified by future experiments. It is known that p-type swCN-FETs can detect analytes through "chemical gating" where a positively (negatively) charged molecular species adsorbs to the nanotube sidewall and locally depletes (enhances) the swCN carrier density leading to a decrease (increase) in current through the FET.[26] This mechanism is consistent with the current decrease and $I(V_G)$ data (Figure 1c) during TMA exposure. Given its pK value of 9.8, TMA should be protonated by residual water (presumably pH ~ 7) associated with the ss-DNA, leading to a rigid shift of the $I(V_G)$ characteristic to the left and a decrease in the sensor current, as observed. The data in Figure 1c correspond to a hole density decrease of ~ 1200/μm due to TMA exposure. Similarly, propionic acid is expected to donate a proton to residual water, in agreement with the measured increase in sensor current. DMMP, along with other chemical nerve agents, is known to be a strong electron donor,[6] consistent with the observed sensor response ($\Delta I/I < 0$). The detection mechanisms for methanol and DNT are less clear. Simple acid-base considerations suggest methanol is neutral under the experimental conditions, and DNT is expected to be an electron acceptor, inconsistent with the measured current decrease for both odors. More detailed experiments are needed to determine whether these species transfer charge to the swCN in the presence of ss-DNA or if detection occurs through a different mechanism, e.g. a conformational change of the ss-DNA that is transduced into an electrical signal by the swCN-FET.

"Electronic nose" detectors are inspired by olfactory systems in biological organisms that typically utilize a thousand different odor receptors, each responsive to many different odorants, to perform amazing feats of molecular identification and analysis. Ss-DNA/swCN-FET gas sensors have a number of properties making them ideal for this application. They are all-electronic sensors with high sensitivity and fast response times (seconds) that compare favorably with well-established and more recently demonstrated sensor families.[27] They offer the advantages of a smaller footprint and simpler implementation than chemicapacitors, and more direct readout with simpler equipment than sensors where odor detection is converted into an optical signal. We demonstrated that the ss-DNA chemical recognition layer is reusable through at least 50 cycles without refreshing or regeneration. It is likely this hybrid nanoscale sensor can be used for liquid phase detection, making it equally appropriate for application in an "electronic tongue" system.[28]

Finally, the intrinsic chemical versatility of ss-DNA, progress in nucleic acid engineering, and use of high-throughput screening may well enable selection of appropriate sequences for detection of a large number of chemical and biological targets. The range of possible targets may be limited by the fact that the ss-DNA chemical recognition component most likely assumes a range of conformations. Future experiments will explore the effectiveness of this sensor class for detection of analytes beyond the small molecules demonstrated here. It has been suggested[29] that an array of 100 sensors with different response characteristics and an appropriate pattern recognition algorithm are sufficient to detect and identify a weak known odor in the face of a strong and variable background. The results presented here represent significant progress towards the realization of a large and diverse sensor array for electronic olfaction, and may bring a practical device within reach when combined with recent progress in fabricating multiplexed arrays of swCN sensors.

**Acknowledgement.** This work was supported by the Laboratory for Research on the Structure of Matter (NSF DMR00-79909) and by the US Department of Energy, grant No. DE-FG02-98ER45701 (M.C.). We thank Michael F. Stern and Dr. Karen McAllister for useful discussions and Dr. Douglas R. Strachan for assistance with data acquisition.